\documentclass[12pt,english]{article}
\usepackage[T1]{fontenc}
\usepackage[latin1]{inputenc}
\usepackage{geometry}
\usepackage{graphicx}
\usepackage{setspace}
\usepackage{amssymb}

\makeatletter

\usepackage{babel}
\makeatother
\begin{document}

\title{Gene Copy Number and Cell Cycle Arrest}

\author{Bhaswar Ghosh and Indrani Bose{*}}

\maketitle
\begin{singlespace}
\begin{center}Department of Physics \end{center}

\begin{center}Bose Institute \end{center}

\begin{center}93/1, A. P. C. Road \end{center}

\begin{center}Kolkata - 700 009, India\end{center}
\end{singlespace}

{*}Author to be contacted for correspondence; e-mail: indrani@bosemain.boseinst.ac.in

Communicated to \emph{Physical Biology}

\begin{abstract}
The cell cycle is an orderly sequence of events which ultimately lead
to the division of a single cell into two daughter cells. In the case
of DNA damage by radiation or chemicals, the damage checkpoints in
the $G_{1}$ and $G_{2}$ phases of the cell cycle are activated.
This results in an arrest of the cell cycle so that the DNA damage
can be repaired. Once this is done, the cell continues with its usual
cycle of activity. We study a mathematical model of the DNA damage
checkpoint in the $G_{2}$ phase which arrests the transition from
the $G_{2}$ to the $M$ (mitotic) phase of the cell cycle. The tumor
suppressor protein p53 plays a key role in activating the pathways
leading to cell cycle arrest in mammalian systems. If the DNA damage
is severe, the p53 proteins activate other pathways which bring about
apoptosis, i.e., programmed cell death. Loss of the p53 gene results
in the proliferation of cells containing damaged DNA, i.e., in the
growth of tumors which may ultimately become cancerous. There is some
recent experimental evidence which suggests that the mutation of a
single copy of the p53 gene (in the normal cell each gene has two
identical copies) is sufficient to trigger the formation of tumors.
We study the effect of reducing the gene copy number of the p53 and
two other genes on cell cycle arrest and obtain results consistent
with experimental observations. 
\end{abstract}

\section*{1. Introduction}

The cell cycle involves the essential mechanism by which a single
cell divides into two daughter cells. Cell division provides the basis
for the growth and development of complex organisms and is required
to replace cells that die with new cells. The cell cycle is a sequence
of orderly events during which a growing cell duplicates all its components
so that after cell division each daughter cell has the same cellular
constituents as the parent cell. In an eukaryotic organism, the cell
cycle progresses through four distinct phases: $G_{1}$, S, $G_{2}$
and M (mitotic) phases {[}1{]}. During the S phase, the chromosomes
containing DNA molecules are replicated. During the M phase, the sister
chromatids are separated and cell division takes place. The $G_{1}$
and $G_{2}$ phases are the gap phases in which cell growth and doubling
of the cellular constituents, other than DNA, take place. Cellular
growth is necessary so that the daughter cells acquire the same size
as that of the parent cell. The cell is arrested at checkpoints if
the processes occurring in the earlier stages of the cycle remain
incomplete. In the case of DNA damage by radiation or chemicals, the
cell cycle is arrested at the damage checkpoints in $G_{1}$and $G_{2}$.
The delay is needed for the repair of the damaged DNA. Once this is
done, progression through the cell cycle is resumed once more. 

The eukaryotic cell operates through the sequential activation and
deactivation of cyclin-dependent protein kinases (CDKs) {[}1{]}. Protein
kinases are enzymes which regulate the structure and/or activity of
target proteins by the transfer of phosphate molecules (phosphorylation).
Similarly, protein phosphatases regulate biochemical activity by removing
phosphate molecules from target proteins (dephosphorylation). CDK
activation can occur only after a regulatory protein cyclin binds
CDK and the complex is phosphorylated by CDK-activating kinases (CAKs).
Even if these conditions are met, the CDK may be inactivated by inhibitory
phosphorylations carried out by certain protein kinases. Protein phosphatases
remove the inhibitory phosphate molecules so that the activity of
the cyclin-CDK complex is triggered. A cycle of synthesis and degradation
of cyclins in each cell cycle controls the periodic assembly and activation
of the cyclin-CDK complexes. In higher eukaryotic organisms, different
cyclin-CDK complexes initiate different cell cycle events. The cell
control system, however, operates on similar principles which may
thus be assumed to be universal. In this paper, we focus attention
on the $G_{2}/M$ transition in frog egg and mammalian cell cycles.
In both the cases, the kinase Wee1 is responsible for the inhibitory
phosphorylation of the cyclin-CDK complex and the phosphatase Cdc25
removes the inhibitory phosphate groups {[}2{]}. The active cyclin-CDK
complex phosphorylates key intracellular proteins which in turn initiate
or control important cell cycle events. The network of molecular interactions,
controlling the stability and activity of the cyclin-CDK complex is
characterized by the presence of feedback loops {[}3{]}. The loops
originate as the proteins (the kinases, phosphatases etc.) influencing
the activity of the cyclin-CDK complex are in turn regulated by the
activity of the cyclin-CDK complex itself. Mathematical models of
the cell cycle have established that the $G_{2}/M$ transition is
analogous to a bistable switch {[}1,2,3,4,5{]}. The bistability arises
because the cyclin-CDK complex inactivates its antagonist the kinase
Wee1 and activates its friend Cdc25C, giving rise to positive feedback
loops. The two stable steady states belong to the $G_{2}$ (low activity
of the cyclin-CDK complex) and the $M$ (high activity of the cyclin-CDK
complex) phases respectively. The transition between the two stable
steady states is not reversible but is described by a hysteresis loop.
The transition from the lower to the upper state occurs when the cyclin
threshold crosses a critical value. Since the amount of cyclin is
correlated with the cell size or cell mass/DNA, the latter quantity
can be treated as the parameter the changing of which triggers the
$G_{2}/M$ transition. Experimental evidence for hysteresis has recently
obtained in a frog egg extract confirming earlier theoretical predictions
{[}6{]}.

We now discuss the function of the DNA damage checkpoint in $G_{2}$
which prevents the entry into the $M$ phase when DNAs are damaged
{[}1,7{]}. The damage induced signal to a series of protein kinases
leads to the phosphorylation and inactivation of the phosphatase Cdc25C.
Dephosphorylation and consequent activation of the cyclin-CDK complex
by Cdc25C are thus inhibited blocking the transition into mitosis.
After the damage is repaired, the DNA damage induced signal is turned
off so that the $G_{2}/M$ transition is possible. In terms of the
hysteresis loop, the cell cycle arrest corresponds to a higher critical
value of the parameter (the cell mass/DNA or analogously the cyclin
concentration) at which the $G_{2}/M$ transition takes place. The
delay in the transition implies an arrest of the cell cycle progression.
In mammalian cells, DNA damage leads to the activation of the tumor
suppressor protein p53 which stimulates the transcription of many
other genes. One of these genes synthesizes the protein p21 which
binds to the cyclin-CDK complex and inhibits its activity. Similarly,
DNA damage activates the chk1 protein which has an inhibiting effect
on Cdc25c activity through phosphorylation. All these processes give
rise to cell cycle arrest allowing for the repair of the DNA damage.
If the p53 gene is mutated, there is a reduction in the amount of
proteins synthesized so that an arrest of the cell cycle may not occur.
As a consequence, the cells containing damaged DNA proliferate through
successive rounds of cell division giving rise to the formation and
growth of tumors. Such cells pose a threat to the organism as they
have a greater probability of becoming cancerous {[}7{]}. In fact,
many cases of human cancer are attributed to mutations in the p53
gene. The p53 protein, by preventing the multiplication of damaged
or stressed cells, acts as a break on the tumor development, hence
the name tumor suppressor. Knudson's well-known two-hit model of tumorigenesis
suggests that mutation of both the copies (diploid organisms have
two copies of each gene) of a tumor suppressor gene is essential for
triggering tumor formation {[}8,9,10{]}. Recent studies, however,
show that the mutation of a single copy is sufficient in many cases
for the loss of the tumor suppression function of the p53 protein.
The gene dosage effect is called haploinsufficiency (HI) and has been
verified experimentally {[}11,12{]}. Tumors are found to arise in
mice with only one intact copy of the p53 gene contrary to Knudson's
hypothesis.

In this paper, we propose a mathematical model of the DNA damage checkpoint
in the $G_{2}$ phase of the cell cycle to illustrate the effect of
gene copy number, i.e., gene dosage on cell cycle arrest. On DNA damage,
the $G_{2}/M$ transition is shown to be arrested when the gene copy
number is two. When the gene copy number is one, the $G_{2}/M$ transition
is not arrested, an effect of HI.

\section*{2. Model of the $G_{2}/M$ transition}

\begin{figure}
\begin{center}\includegraphics[%
  width=4in]{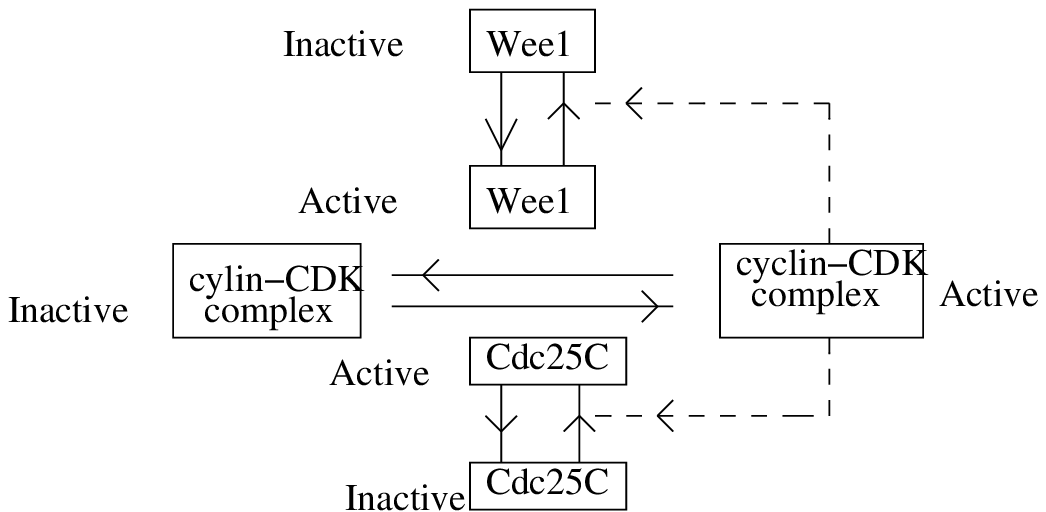}\end{center}

Figure 1. Molecular interactions of the cyclin-CDK complex with Wee1
and Cdc25C
\end{figure}

We consider a mammalian system in which the kinase Wee1 inhibits the
activity of the cyclin-CDK complex and the phosphatase Cdc25C activates
the complex through dephosphorylation. On the other hand, the cyclin-CDK
complex in its active state acts as a kinase and inactivates Wee1
through phosphorylation. The complex further activates Cdc25C via
phosphorylation. The molecular interactions of the cyclin-CDK complex
with Wee1 and Cdc25C are shown in figure 1. The phosphate groups are
not shown in the figure. The interactions of the cyclin-CDK complex
with Wee1 and Cdc25C constitute two positive feedback loops shown
in figure 2. The arrow sign denotes activation and the $\perp$ sign
denotes inhibition. The general reaction schemes for phosphorylation
and dephosphorylation reactions are:

\begin{equation}
S_{1}+E_{1}\quad\begin{array}{c}
k_{1}\\
\rightleftharpoons\\
k_{2}\end{array}\quad E_{1-}S_{1}\quad\begin{array}[b]{c}
k_{3}\\
\longrightarrow\end{array}\quad S_{2}+E_{1}\label{eq:1}\end{equation}

\begin{equation}
S_{2}+E_{2}\quad\begin{array}{c}
k_{1}^{'}\\
\rightleftharpoons\\
k_{2}^{'}\end{array}\quad E_{2-}S_{2}\quad\begin{array}[b]{c}
k_{3}^{'}\\
\longrightarrow\end{array}\quad S_{1}+E_{2}\label{eq:2}\end{equation}

\begin{figure}
\begin{center}\includegraphics[%
  width=4in]{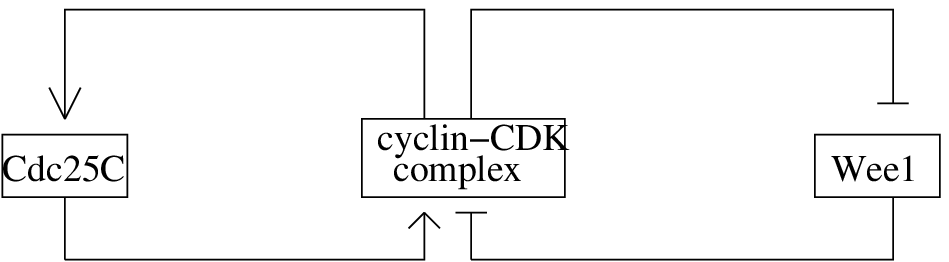}\end{center}

Figure 2. Two positive feedback loops involving the cyclin-CDK complex,
Wee1 and Cdc25C 
\end{figure}

\noindent In equations (1) and (2), $S_{1}$ and $S_{2}$ denote
the dephosphorylated and phosphorylated proteins respectively, $E_{1}$
($E_{2}$) denotes the kinase (phosphatase) catalyzing the phosphorylation
(dephosphorylation) reaction. Assuming Michaelis-Menten reaction kinetics,
the differential equations describing the rates of change of $S_{1}$
and $S_{2}$ w.r.t. time are given by

\begin{equation}
\frac{ds{}_{1}}{dt}=j-k_{3}\: E_{1T}\:\frac{s_{1}}{j_{1}+s_{1}}+k_{3}^{'}\: E_{2T}\:\frac{s_{2}}{j_{2}+s_{2}}-\gamma_{1}s_{1}\label{eq:3}\end{equation}

\begin{equation}
\frac{ds{}_{2}}{dt}=k_{3}\: E_{1T}\:\frac{s_{1}}{j_{1}+s_{1}}-k_{3}^{'}\: E_{2T}\:\frac{s_{2}}{j_{2}+s_{2}}-\gamma_{2}s_{2}\label{eq:4}\end{equation}

\noindent In equation (3), $s_{1}$ and $s_{2}$ denote the concentrations
of the dephosphorylated and phosphorylated proteins respectively.
Synthesis of $S_{1}$ at a constant rate $j$ and decay of $S_{1}$
are also considered. Equation (4) includes the decay term for $S_{2}$.
$E_{1T}$ and $E_{2T}$ are the total amounts of the kinase and the
phosphatase and $j_{1}$, $j_{2}$ are the Michaelis-Menten constants.
The interactions shown in figure 1 describe phosphorylation and dephosphorylation
reactions. Reaction schemes in (1) and (2) and the associated differential
equations in (3) and (4) lead to the following set of differential
equations describing the network shown in figure 1: 

\begin{equation}
\frac{dx_{1}}{dt}=j_{c}m-k_{1}\: y\:\frac{x_{1}}{j_{1}+x_{1}}+k\: z\:\frac{x_{2}}{j_{2}+x_{2}}-\gamma_{c1}x_{1}\label{eq:5}\end{equation}

\begin{equation}
\frac{dx_{2}}{dt}=k_{1}\: y\:\frac{x_{1}}{j_{1}+x_{1}}-k\: z\:\frac{x_{2}}{j_{2}+x_{2}}-\gamma_{c2}x_{2}\label{eq:6}\end{equation}

\begin{equation}
\frac{dy}{dt}=k_{2}\:\frac{1-y}{j_{3}+1-y}-k_{2}^{'}\: x_{1}\:\frac{y}{j_{4}+y}\label{eq:7}\end{equation}

\begin{equation}
\frac{dz}{dt}=k_{3}\: x_{1}\;\frac{z_{1}}{j_{5}+z_{1}}-k_{3}^{'}\:\frac{z}{j_{6}+z}-\beta_{1}\: z\label{eq:8}\end{equation}

\begin{equation}
\frac{dz_{1}}{dt}=\alpha-k_{3}\: x_{1}\;\frac{z_{1}}{j_{5}+z_{1}}+k_{3}^{'}\:\frac{z}{j_{6}+z}-\beta_{1}\: z_{1}\label{eq:9}\end{equation}

\noindent where $x_{1}$ ($x_{2}$) denotes the concentration of
the active (inactive) cyclin-CDK complex, $y$ is the concentration
of the active Wee1 and $z$ ($z_{1}$) denotes the concentration of
the active (inactive) Cdc25C. The total concentration of Wee1 is normalized
to one so that $1-y$ represents the concentration of the inactive
Wee1. The $j_{i}$'s ($i=1,....,6$) are the Michaelis-Menten constants.
The first term in equation (5) arises due to the synthesis of cyclins,
the rate of which is proportional to the cell mass/DNA m. In the $G_{2}$
phase, cells are growing and larger cells, it is assumed, synthesize
cyclin at a higher rate {[}13{]}. The steady state solutions of equations
(5)-(9) are obtained with the help of Mathematica. Figure 3 shows
the result in the form of a hysteresis loop with $m$ playing the
role of the bifurcation parameter. In the region of bistability, the
two stable steady states correspond to the $G_{2}$ (lower branch)
and $M$ phases respectively. When $m$ and correspondingly the cyclin
concentration in the cyclin-CDK complex reaches a threshold value,
the active complex inactivates Wee1 and activates Cdc25C in sufficient
amounts. This triggers the autocatalytic conversion of the inactive
cyclin-CDK complex into the active complex and a transition to the
mitotic phase, with a higher concentration of the active cyclin-CDK
complex, takes place. Results shown in figure 3 have been obtain for
the following parameter values. The rate constants are (in units of
$min^{-1}$)

\noindent$j_{c}=0.02,$ $k=0.01,$ $k_{1}=0.01,$ $\gamma_{c1}=0.01,$
$\gamma_{c2}=0.01,$ $k_{2}=0.004,$ $k_{2}^{'}=0.2,$ $k_{3}=0.004,$
$k_{3}^{'}=0.001$, $\alpha=0.01,$ $\beta_{1}=0.002.$

\noindent The Michaelis-Menten constants (dimensionless) are:

\noindent$j_{1}=0.5,$ $j_{2}=0.5,$ $j_{3}=0.02,$ $j_{4}=0.02,$
$j_{5}=0.02,$ $j_{6}=0.02$. 

\noindent Values of the different rate and Michaelis-Menten constants
are consistent with those reported in earlier literature {[}4,5,13{]}
and the forms of the differential equations (equations (5)-(8)) are
similar. Bistability and the hysteresis loop are obtained for a wide
range of parameter values and changes in the structure of the differential
equations, like replacing the Michaelis-Menten kinetics by the mass
action kinetics, do not alter the basic result. In the next section,
we propose a mathematical model of the $G_{2}$ damage checkpoint
which, combined with the model of this section, predicts a cell cycle
arrest on DNA damage. Progression of the cell cycle, however, is not
halted if the p53 gene copy number is one instead of two.

\begin{figure}
\begin{center}\includegraphics[%
  width=3in]{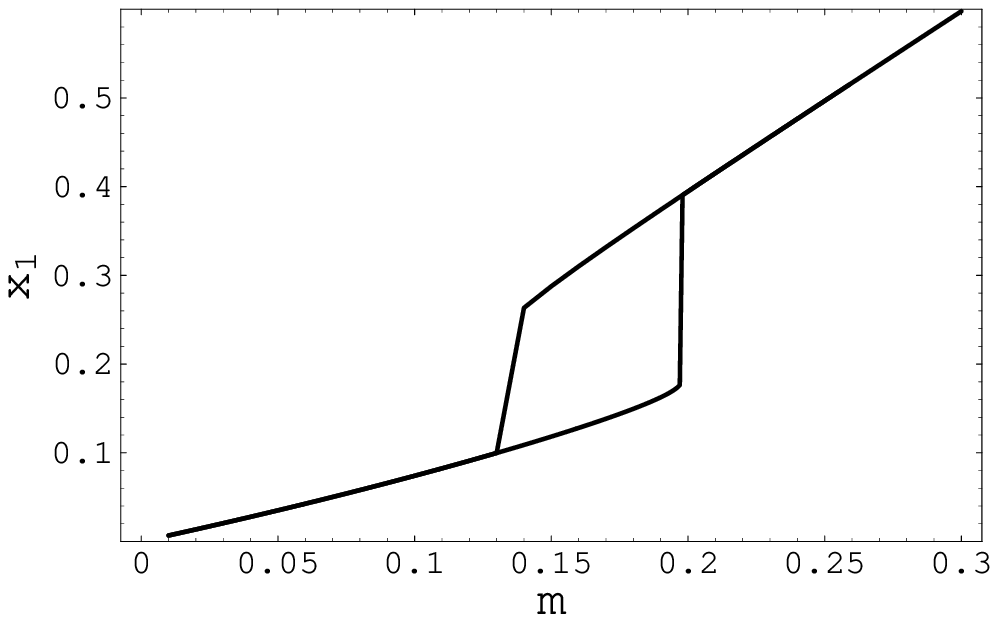}\end{center}

Figure 3. Hysteresis loop for the $G_{2}/M$ transition; $x_{1}$
is the concentration of the active cyclin-CDK complex and $m$ the
cellular mass/DNA
\end{figure}

\section*{3. DNA damage and cell cycle arrest}

We now discuss how the p53 network is activated on DNA damage in the
$G_{2}$ phase of the cell cycle. In a normal cell, the level of the
p53 protein is low {[}7{]}. The amount depends mostly on the degradation
rate of the protein rather than on its rate of synthesis. The p53
protein stimulates the transcription of the MDM2 gene and the MDM2
protein binding to the p53 protein activates its degradation through
ubiquitin-mediated proteolysis. The interactions between the p53 and
the MDM2 proteins give rise to a negative feedback loop which keeps
the p53 protein level low. On DNA damage, a protein called ATM kinase
is activated which phosphorylates the p53 protein at a specific site
and prevents the binding of the MDM2 protein to the p53. This inhibits
the MDM2 mediated degradation of the p53 so that the protein stabilizes
at a higher level. The protein p53 activates the transcription of
the MDM2 gene giving rise to a negative feedback loop. An effective
positive feedback loop coexists with the negative feedback loop (figure
4) due to the p53 mediated inhibition of the transport of the cytoplasmic
MDM2 proteins into the cell nucleus {[}14{]}. Experimental observations
suggest that the p53 and MDM2 proteins are degraded mainly in the
nucleus {[}15{]}. There is also experimental evidence that DNA damage
kinases induce an auto-degradation of the MDM2 proteins {[}15{]}.
Recently, Lahav et al. {[}16{]} have performed a single cell experiment
on the dynamics of the p53-MDM2 network after DNA damage. The response
of the network to the damage is found to be digital in the form of
a discrete number of p53 and MDM2 protein pulses. The amplitude of
the pulses and the interpulse time interval do not depend on the amount
of DNA damage but the number of pulses is determined by the magnitude
of the damage. A negative feedback loop can give rise to oscillations
if the number of elements in the loop exceeds two or a time delay
is included in the feedback process. A simple mathematical model of
the p53-MDM2 negative feedback loop includes an intermediary of unknown
origin in order to obtain oscillations {[}17{]}. A more recent study
{[}14{]} proposes a mathematical model in which the p53-MDM2 interaction
network is based on both negative and positive feedbacks. The model
considers in detail the mechanisms which contribute to the two feedback
loops. The intermediate processes included in the negative feedback
loop introduce the time delay required for getting oscillations. In
our simplified model, we assume the existence of feedback loops as
shown in figure 4 without explicitly taking into account the biochemical
events contributing to the repression of MDM2 by p53. Oscillations
in our simplified scheme are generated by considering the expression
of the MDM2 gene to be a two-step process, i.e, consisting of both
transcription and translation. In the earlier modeling studies {[}14,17{]},
only one step, namely, protein synthesis was taken into account. The
two-step GE introduces a time delay in the p53-MDM2 network dynamics
leading to oscillations. The p53 pulses further activate the transcription
of a gene p21. The p21 proteins inhibit the activity of the cyclin-CDK
complex leading to a delay in the $G_{2}/M$ transition, i.e., a cell
cycle arrest {[}7{]}. As mentioned in the introduction, DNA damage
also activates the chk1 protein which inhibits Cdc25C activity. Figure
5 shows the DNA damage response network which arrests the $G_{2}/M$
transition. In figure 5, the p53 protein activity is determined by
the processes depicted in figure 4. Similarly the cyclin-CDK complex
and Cdc25C are parts of the network shown in figure 1. The differential
equations describing the levels of p53 and MDM2 activity (figures
4 and 5) are:

\begin{equation}
\frac{dp}{dt}=j_{0}+x_{d}\:\frac{p_{a}}{j_{1}^{'}+p{}_{a}}-A\:\frac{p}{j_{2}^{'}+p}-\beta_{2}\,\, p\: m_{p}-\beta_{1}\: p\label{eq:10}\end{equation}

\begin{equation}
\frac{dp_{a}}{dt}=-x_{d}\:\frac{p_{a}}{j_{1}^{'}+p{}_{a}}+A\:\frac{p}{j_{2}^{'}+p}-\beta_{1}\: p\label{eq:10}\end{equation}

\begin{equation}
\frac{dm_{r}}{dt}=s_{m}+j_{m}\:\frac{k_{a}^{'}}{k_{a}^{'}+k_{d}^{'}}-\gamma_{m}\: m_{r}\label{eq:11}\end{equation}

\begin{equation}
\frac{dm_{p}}{dt}=j_{p}\: m_{r}-\gamma_{p}\: m_{p}-k_{1m}\:(p+p_{a})\: m_{p}-k_{2m}\: A\: m_{p}\label{eq:12}\end{equation}

\begin{equation}
\frac{dm21}{dt}=j_{21}\:\frac{k_{a}^{''}}{k_{a}^{''}+k_{d}^{''}}-\gamma_{m}\: m21\label{eq:13}\end{equation}

\begin{equation}
\frac{dp21}{dt}=j_{p21}\: m21-\gamma_{p21}\: p21\label{eq:14}\end{equation}

\noindent with

\begin{equation}
k_{a}^{'}=k_{a}\;\frac{\{(p+p_{a})/k_{m}\}^{4}}{1+\{(p+p_{a})/k_{m}\}^{4}},\qquad k_{d}^{'}=k_{d}\label{eq:15}\end{equation}

\begin{equation}
k_{a}^{''}=k_{a}\;\frac{(p_{a}/k_{21})^{4}}{1+(p_{a}/k_{21}){}^{4}},\qquad k_{d}^{''}=k_{d}\label{eq:16}\end{equation}

\begin{figure}
\begin{center}\includegraphics[%
  width=4in]{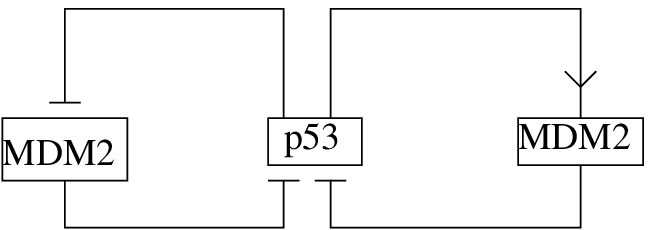}\end{center}

Figure 4. The p53-MDM2 network describing the effective interactions
between the p53 and the MDM2 proteins
\end{figure}

\noindent In the equations, $p$ and $p_{a}$ denote the concentrations
of the inactive and the active p53, $m_{r}$ and $m_{p}$ are the
concentrations of the MDM2 mRNA and protein, $m21$ and $p21$ represent
the concentrations of the p21 mRNA and protein and $A$ denotes the
amount of DNA damage. In equation (10), the first term describes the
production of the p53 protein (inactive), the second term corresponds
to the dephosphorylation, i.e., inactivation of the active p53, the
third term refers to the phosphorylation, i.e., activation of the
p53 on DNA damage, the fourth term describes the p53 degradation by
the MDM2 protein and the last term refers to the p53 degradation.
Equation (12) describes the synthesis of the MDM2 mRNA, the first
term of which represents the basal rate of production. The second
term denotes mRNA production due to the activation of the MDM2 gene
by both the active and inactive forms of p53. The MDM2 gene can be
in two states: inactive and active. In the inactive state of the gene,
mRNA production takes place at the basal rate $s_{m}$. In the active
state of the gene (activation brought about by the transcription factor
p53), mRNA production occurs with rate constant $j_{m}$. The p53
proteins tetramerize to regulate the MDM2 gene expression. The reaction
scheme describing the production and degradation of mRNA and protein
is given by

\begin{equation}
\begin{array}{c}
G+S\quad\begin{array}{c}
\\\rightleftharpoons\\
k_{m}\end{array}\quad G_{-}S\quad\begin{array}{c}
k_{a}\\
\rightleftharpoons\\
k_{d}\end{array}\quad G^{\star}\quad\begin{array}[b]{c}
j_{m}\\
\longrightarrow\end{array}\quad m_{r}\quad\begin{array}[b]{c}
\gamma_{m}\\
\longrightarrow\end{array}\quad\Phi\\
G\quad\begin{array}[b]{c}
s_{m}\\
\longrightarrow\end{array}\quad m_{r}\quad\begin{array}[b]{c}
\gamma_{m}\\
\longrightarrow\end{array}\quad\Phi\\
m_{r}\quad\begin{array}[b]{c}
j_{p}\\
\longrightarrow\end{array}\quad m_{p}\quad\begin{array}[b]{c}
\gamma_{p}\\
\longrightarrow\end{array}\quad\Phi\end{array}\label{eq:}\end{equation}

\begin{figure}
\begin{center}\includegraphics[%
  width=2in]{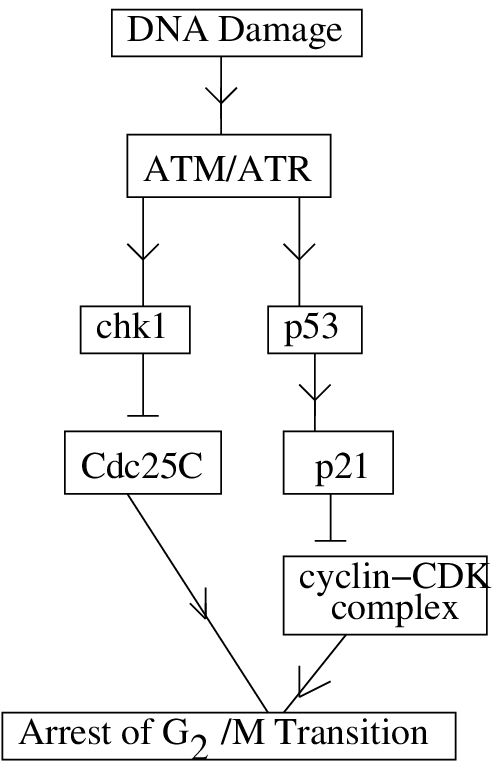}\end{center}

Figure 5. DNA damage response network leading to cell cycle arrest
\end{figure}

\noindent where $G$ ($G^{*}$) denotes the inactive (active) state
of the gene, $S$ denotes the p53 tetramer, $k_{m}$ is the equilibrium
dissociation constant, $\gamma_{m}$, $\gamma_{p}$ the rate constants
for mRNA and protein degradation and $k_{a}$, $k_{d}$, the activation
and deactivation rate constants. In equation (11), $k_{a}^{'}$ and
$k_{d}^{'}$ denote the effective rate constants for activation and
deactivation and are given in equation (16). In equation (13), the
third term takes into account the inhibiting effect of p53 on the
MDM2 protein (figure 4) while the fourth term denotes the damage induced
degradation of the MDM2 protein {[}15{]}. In equation (14), the first
term arises due to the activation of the p21 gene by the activated
p53 protein. The effective rate constants $k_{a}^{''}$ and $k_{d}^{''}$
are analogous to the rate constants $k_{a}^{'}$ and $k_{d}^{'}$
in equation (11) and are given in equation (17). Figure 6 shows the
response of the p53 network (figures 4 and 5) to DNA damage. The p53
protein is produced in pulses consistent with experimental observation
{[}16{]}. The rate constants are (in units of $min^{-1}$) 

\begin{figure}
\begin{center}\includegraphics[%
  width=3in]{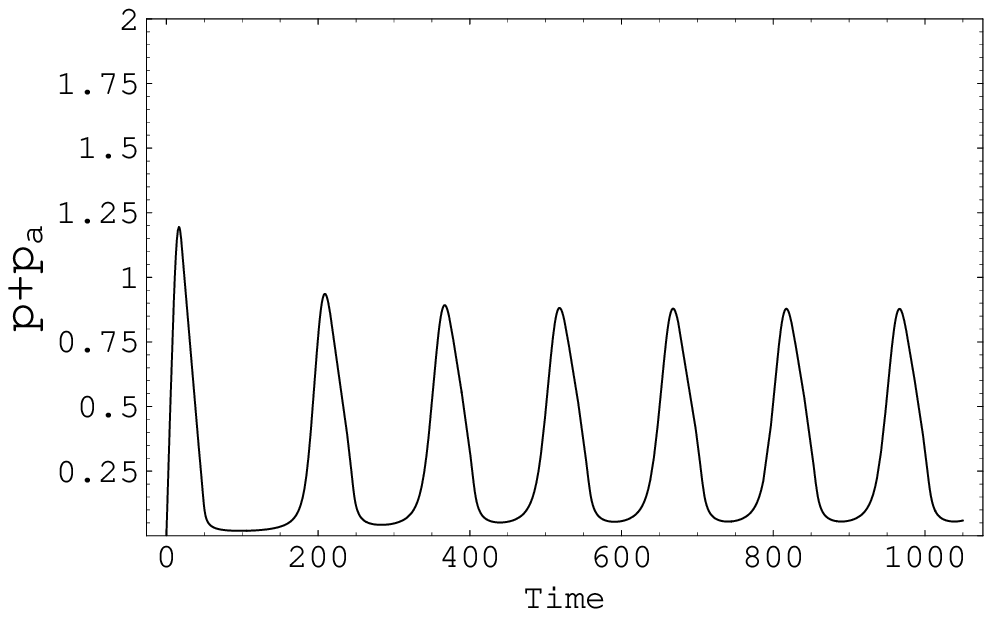}\end{center}

Figure 6. Pulses of p53 proteins generated on DNA damage; $(p+p_{a}$)
is the total p53 concentration.
\end{figure}

\noindent$j_{0}=0.1$, $x_{d}=0.04$, $\beta_{2}=0.2$, $\beta_{1}=0.01$,
$s_{m}=0.0001$, $j_{m}=1$, $k_{a}=10$, $k_{d}=10$, $\gamma_{m}=0.01$,
$j_{p}=2$, $\gamma_{p}=0.01$, $k_{1m}=2$, $k_{2m}=0.01$, $j_{21}=0.4$,
$j_{p21}=2$, $\gamma_{p21}=0.005$

\noindent The Michaelis-Menten constants (dimensionless) are

\noindent$j_{1}^{'}=0.01$, $j_{2}^{'}=2$.

\noindent The binding constants $k_{m}$ and $k_{21}$ are 

\noindent $k_{m}=2$, $k_{21}=1$. The amount of DNA damage is $A=0.2$.

We now study the cell cycle arrest on DNA damage by combining the
operations of the cyclin-CDK complex (figure 1) and the DNA damage
response network (figures 4 and 5 together). Since the chk1 proteins
inhibit the activity of Cdc25C and the p21 proteins inactivate the
active cyclin-CDK complex, the inclusion of these processes in equations
(5) and (8) leads to the following modified equations:

\begin{equation}
\frac{dx_{1}}{dt}=j_{c}m-k_{1}\: y\:\frac{x_{1}}{j_{1}+x_{1}}+k\: z\:\frac{x_{2}}{j_{2}+x_{2}}-\gamma_{c1}x_{1}-\delta\: p21\: x_{1}\label{eq:18}\end{equation}

\begin{equation}
\frac{dz}{dt}=k_{3}\: x_{1}\;\frac{z_{1}}{j_{5}+z_{1}}-k_{3}^{'}\:\frac{z}{j_{6}+z}-\beta_{1}\: z-k_{4}\: p_{chk1}^{a}\: z\label{eq:19}\end{equation}

\begin{figure}
\begin{center}\includegraphics[%
  width=3in]{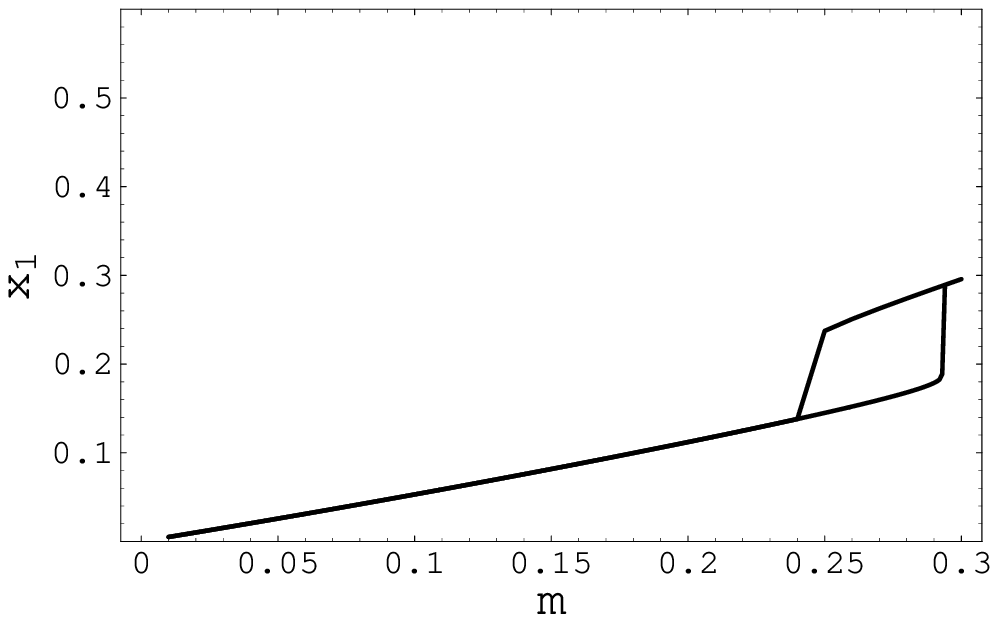}\end{center}

Figure 7. Cell cycle arrest on DNA damage; $x_{1}$ is the concentration
of the active cyclin-CDK complex and $m$ the cellular mass/DNA
\end{figure}

\noindent The last terms in equations (19) and (20) take into account
the damage induced interactions mentioned above. The time evolution
of the concentrations of the inactive ($p_{chk1}^{in}$) and the active
($p_{chk1}^{a}$) chk1 proteins is given by

\begin{equation}
\frac{dp_{chk1}^{in}}{dt}=j_{chk1}+\omega_{1}\:\:\frac{p_{chk1}^{a}}{j_{3}^{'}+p_{chk1}^{a}}-\omega_{2}\: A\:\frac{p_{chk1}^{in}}{j_{4}^{'}+p_{chk1}^{in}}-\gamma_{chk1}\: p_{chk1}^{in}\label{eq:20}\end{equation}

\begin{equation}
\frac{dp_{chk1}^{a}}{dt}=-\omega_{1}\:\:\frac{p_{chk1}^{a}}{j_{3}^{'}+p_{chk1}^{a}}+\omega_{2}\: A\:\frac{p_{chk1}^{in}}{j_{4}^{'}+p_{chk1}^{in}}-\gamma_{chk1}\: p_{chk1}^{a}\label{eq:}\end{equation}

\noindent Figure 7 is obtained for $\delta=0.004$, $k_{4}=0.1$,
$j_{chk1}=0.01$ $\omega_{1}=0.1$, $\omega_{2}=2$, $j_{3}^{'}=0.01$,
$j_{4}^{'}=0.03$ and $\gamma_{chk1}=0.01$ by solving the set of
differential equations (6), (7), (9), (19), (20), (21), (22) and (10)-(15).
The values of the other constants are as already specified. Comparing
figure 7 with figure 3, one finds that the the cell cycle is arrested
due to a shift in the critical value of the cell mass/DNA, required
for the $G_{2}/M$ transition, to a higher value. The delay allows
the repair of DNA damage to be undertaken. Figure 7 has been obtained
assuming the gene copy number of the p53 gene to be two. We now consider
the case when one of the gene copies is mutated. This gives rise to
a fall in the amount of proteins synthesized by 50\%. The effect of
reduced gene copy number is incorporated in the mathematical model
by halving the value of the rate constant $j_{0}$ (equation (9))
associated with the synthesis of p53 proteins. Figure 8 shows the
hysteresis loop when the gene copy number is reduced to one. Comparison
of figures 3 and 8 show that the cell cycle is not arrested and the
$G_{2}/M$ transition occurs at the same critical value of the cell
mass/DNA as in the case of undamaged DNA. 

The failure of the DNA damage response network to arrest the cell
cycle when the gene copy number of the p53 gene is one provides a
clear demonstration of HI associated with the p53 gene. A question
that naturally arises in this context is whether the chk1 and MDM2
genes also exhibit a similar effect. In the case of the chk1 gene,
reduced gene copy number implies a diminished production of $p_{chk1}$
proteins. The effect is taken into account by replacing $j_{chk1}$,
the rate constant for protein synthesis, by $j_{chk1}/2$ in equation
(21). In this case, one finds that the cell cycle is still arrested
with almost the same time delay as in the case when the gene copy
number of the chk1 gene is two. An interaction not included in the
DNA damage response network of figure 5 is the activation of the p53
protein by the activated chk1 protein {[}18{]}. When this interaction
is included, the chk1 exhibits HI, i,e., the cell cycle is not arrested
when the gene copy number of the chk1 gene is reduced from two to
one. In the case of the MDM2 gene, reduction of the gene copy number
to one ($s_{m}$ and $j_{m}$ in equation (12) replaced by $s_{m}/2$
and $j_{m}/2$) introduces a larger time delay, i.e., shifts the $G_{2}/M$
transition point to a higher value of $m$ (figure 9) but the region
of bistability shrinks to a tiny size. With the increase in the amount
of DNA damage the hysteretic transition is lost. The small size of
the region of bistability and its disappearance indicate abnormal
cell cycle function and provide a different manifestation of HI. Figure
10 shows the $G_{2}/M$ hysteretic transition when the gene copy number
of both the p53 and MDM2 genes is one. In this case, one finds that
the cell cycle is arrested though with lesser time delay than in the
case when the gene copy number of both the p53 and MDM2 genes is two.

Mendrysa et al. {[}19{]} have shown that a reduction of the MDM2 protein
level in vivo results in increased radiosensitivity. In our study
of the DNA damage checkpoint in $G_{2}$, we assumed the amount of
DNA damage to be $A=0.2$. We now consider $A$ to have the magnitude
0.17, i.e., reduced DNA damage. Figure 11 shows the $G_{2}/M$ transition
when both the p53 and MDM2 genes have gene copy number two. In the
absence of the DNA damage, the $G_{2}/M$ transition occurs as in
figure 3. Comparison of the figures 3 and 11 show that the time delay
in the $G_{2}/M$ transition is practically negligible, i.e., the
damage response network is insensitive to DNA damage. Figure 12 shows
the same transition when the gene copy number of the MDM2 gene is
reduced to one. This gives rise to decreased MDM2 protein levels and
a time delay in the $G_{2}/M$ transition. In other words, the DNA
damage response network becomes more sensitive with reduced MDM2 protein
levels. With increased DNA damage, diminished MDM2 protein levels
may give rise to abnormal cell cycle features like tiny hysteresis
loops (figure 9) or the absence of a region of bistability. These
examples indicate an increased damage sensitivity with decreased MDM2
protein levels and may sometimes have lethal consequences for the
cellular integrity.

\section*{4. Conclusion and Outlook}

In this paper, we study a mathematical model of the $G_{2}/M$ transition
in the mammalian cell division cycle and discuss how this transition
is arrested when the cellular DNA is damaged by radiation or chemicals.
The $G_{2}/M$ transition is a result of bistability with the two
stable steady states associated with the $G_{2}$ and $M$ phases.
Bistability arises due to the presence of positive feedback loops
in the network involving the cyclin-CDK complex, Wee1 and Cdc25C (figures
1 and 2). Bistability is further accompanied by a hysteresis loop
(figure 3) the existence of which has been verified experimentally
{[}6{]}. The function of the DNA damage response network (figure 5),
when activated by damage induced signals, is to halt the progression
of the cell cycle, i.e., to arrest the $G_{2}/M$ transition. The
arrest introduces a time delay in the transition as a greater amount
of cyclin or equivalently cell mass/DNA is required for the transition
to occur. A core component of the damage response network is the p53-MDM2
network with effective molecular interactions shown in figure 4. In
a normal cell, the p53 protein level is low. On DNA damage, the p53
and MDM2 proteins are produced in pulses (figure 6) consistent with
experimental observation {[}16{]}. The pathways initiated by the activated
p53 and chk1 proteins contribute towards the arrest of the $G_{2}/M$
transition (figure 7).

The full mathematical model is described by the set of differential
equations (6), (7), (9), (19)-(22) and (10)-(15). The equations describing
the network in figure 1 are similar to those in the pioneering studies
of Tyson and collaborators {[}4,5{]}. The genesis of the effective
interactions in the p53-MDM2 network (figure 4) is explained in detail
in Ref. {[}14{]}. We consider the model of figure 4 and take explicit
account of the separate steps of transcription and translation in
the production of the MDM2 proteins. The simplified mathematical framework
is adequate for obtaining p53 and MDM2 oscillations. The modeling
of the chk1 and p53 activated pathways in the damage response network
of figure 5 is a new contribution. Aguda has earlier carried out a
quantitative analysis of the kinetics of the $G_{2}$ DNA damage checkpoint
system {[}20{]}. His model is more detailed than ours but does not
include the experimentally observed features like hysteresis and pulsed
production of the p53 and MDM2 proteins. The major focus of our study
is to study the effect of gene dosage (gene copy number) on the $G_{2}/M$
transition in the case of DNA damage. To our knowledge, such studies
have not been carried out earlier. Borisuk and Tyson {[}21{]} have
addressed the issue of gene dosage briefly vis-$\grave{a}$-vis its
effect on limit cycle oscillations describing periodic cell division
and have concluded that the period of the limit cycle is in general
insensitive to gene dosage (two-fold change in parameter values have
been considered). In a diploid organism, each gene exists in two copies
and when one of these is mutated, the gene copy number is reduced
to one. Consequently, the amount of proteins synthesized is reduced
and may fall below a threshold level for the onset of some desired
activity. This can give rise to HI, a manifestation of which is in
the form of a disease {[}22,23{]}. A large number of diseases are
caused by mutations in genes encoding proteins called transcription
factors (TFs). More than 30 different human maladies have been attributed
to TF HI. TFs regulate GE by binding at the appropriate region of
the DNA. Cooperative interactions among the TFs favour the formation
of bound TF complexes (oligomers). Such multimeric complexes are essential
for the initiation of GE in many eukaryotic systems. The TFs constitute
the stimulus (concentration $S$) and the response is quantified by
the amount of proteins synthesized (concentration $R$) from the target
gene. Due to the multimerization of the TFs, the curve $R$ versus
$S$ has a sigmoidal shape. A small change in $S$ around the inflection
point (the point at which the tangent to the curve has the maximum
slope) gives rise to a significant change in the amount $R$ of the
response. Thus, if there are two TF encoding genes and one of these
is mutated, the level of the TFs produced may fall below the inflection
point of the sigmoid leading to a considerable reduction in the amount
$R$ of the proteins synthesized from the regulated gene. TF HI occurs
if the amount of proteins synthesized falls below a threshold level
for the onset of protein activity. In the DNA damage response network
(figure 5), the activating p53 proteins tetramerize and act as TFs
in regulating the expression of the target genes. Two of these genes
are the MDM2 and the p21 genes. The TF activity of the p53 proteins,
on DNA damage, initiates processes leading to the cell cycle arrest.
The delay in the $G_{2}/M$ transition is desirable to allow time
for the repair of the DNA damage. If repair is not possible, the p53
proteins activate other pathways leading to apoptosis, i.e., programmed
cell death {[}1,7{]}. In the absence of p53 proteins, cell cycle arrest
and apoptosis are not possible leading to a proliferation of cells
(containing damaged DNA) through successive cell divisions. The unchecked
rounds of cell cycle give rise to the formation and growth of tumors
which in many cases turn cancerous. Some current studies {[}9,10{]}
suggest that certain forms of cancer may be a result of TF HI arising
from a reduced gene copy number of the p53 tumor suppressor gene.
The main result of our study is to show that on DNA damage, the cell
cycle is arrested (not arrested) in the $G_{2}$ phase when the gene
copy number of the p53 gene is two (one). This is a clear demonstration
of HI and one can verify that the result continues to hold true for
many other parameter values. We have further explored the issue of
whether the chk1 and MDM2 genes also exhibit HI. Our mathematical
model predicts observable consequences in both the cases when the
gene copy number is reduced to one. The chk1 gene like the p53 gene
is a tumor suppressor. The chk1, a protein kinase, is involved in
transducing the DNA damage signals and is needed for the operation
of the intra-S phase and $G_{2}/M$ checkpoints. Experimental evidence
of HI has been obtained in the first case and the possibility of HI
in the second case is also not ruled out {[}24,25{]}. Our model predicts
increased sensitivity of the DNA damage response network when the
MDM2 protein levels are reduced which is in agreement with experimental
observations {[}19{]}. The proposal that HI is responsible for the
occurrence of some types of cancer is of recent origin. Experimental
evidence on HI related to the DNA damage response of the cell cycle
is now beginning to be accumulated. The molecular mechanisms responsible
for the control of the damage response are in most cases not fully
known. Our preliminary study is meant to highlight the non-trivial
effect of gene copy number on the cell cycle response to DNA damage
in the $G_{2}$ phase of the mammalian cell cycle. Further studies
based on more detailed mathematical models are needed for a clearer
understanding of the problem.

\begin{figure}
\begin{center}\includegraphics[%
  width=3in]{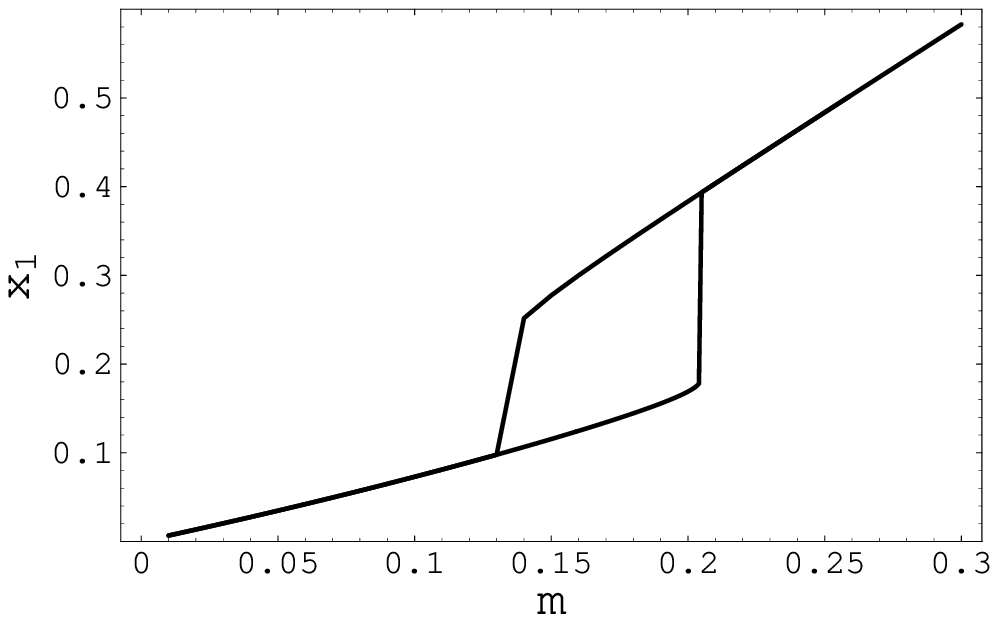}\end{center}

Figure 8. The $G_{2}/M$ transition with one copy of the p53 gene;
$x_{1}$ is the concentration of the active cyclin-CDK complex and
$m$ the cellular mass/DNA
\end{figure}

\begin{figure}
\begin{center}\includegraphics[%
  width=3in]{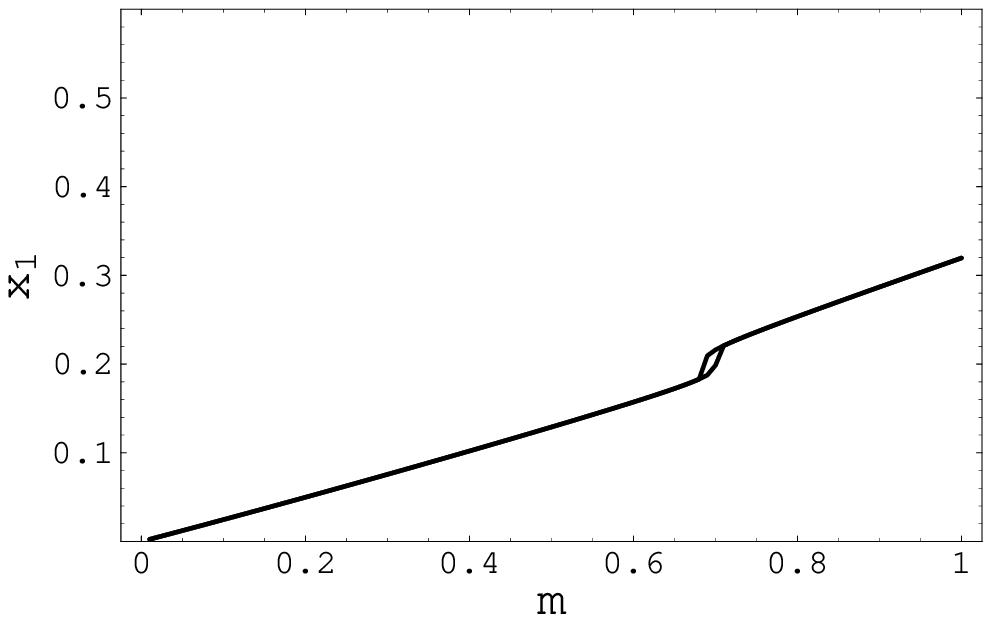}\end{center}

Figure 9. The $G_{2}/M$ transition for one copy of the MDM2 gene
and two copies of the p53 gene; $x_{1}$ is the concentration of the
active cyclin-CDK complex and $m$ the cellular mass/DNA
\end{figure}

\begin{figure}
\begin{center}\includegraphics[%
  width=3in]{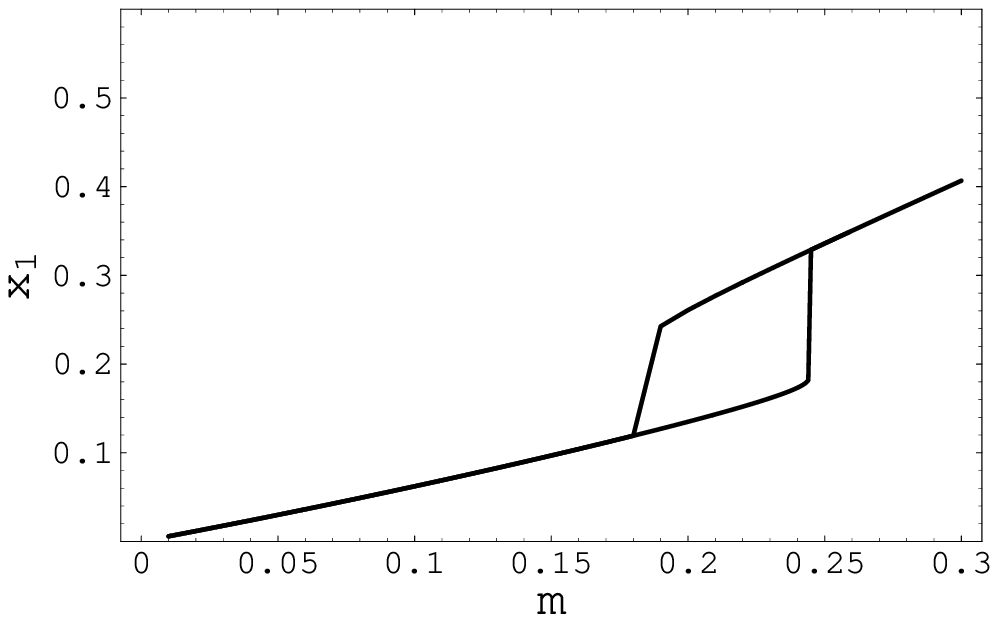}\end{center}

Figure 10. The $G_{2}/M$ transition when the gene copy number of
both the p53 and MDM2 genes is one; $x_{1}$ is the concentration
of the active cyclin-CDK complex and $m$ the cellular mass/DNA
\end{figure}

\begin{figure}
\begin{center}\includegraphics[%
  width=3in]{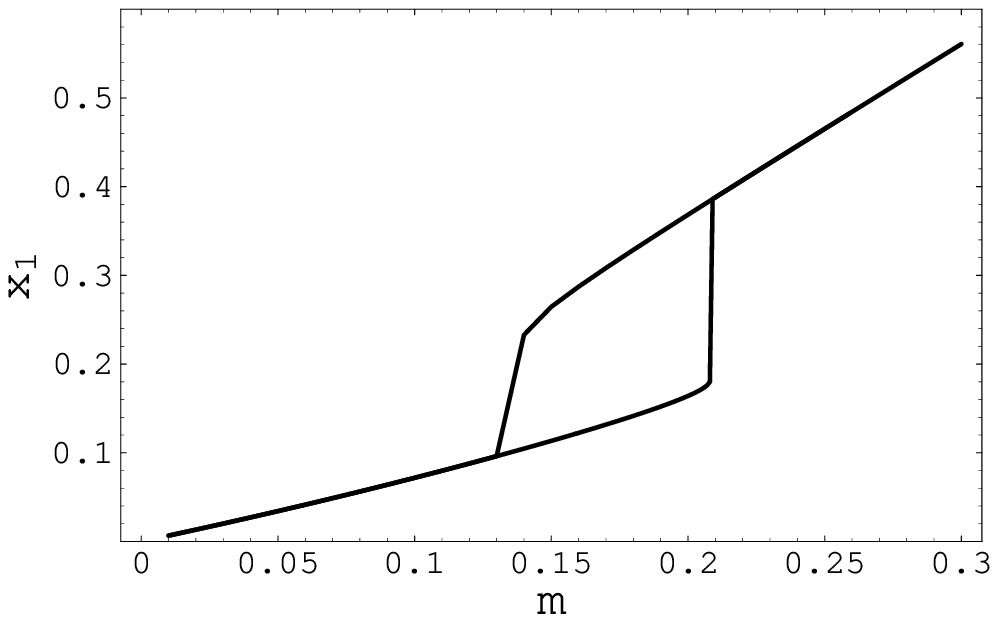}\end{center}

Figure 11. The $G_{2}/M$ transition when the gene copy number of
both the p53 and MDM2 genes is two and the amount of DNA damage is
$A=0.17$. 
\end{figure}

\begin{figure}
\begin{center}\includegraphics[%
  width=3in]{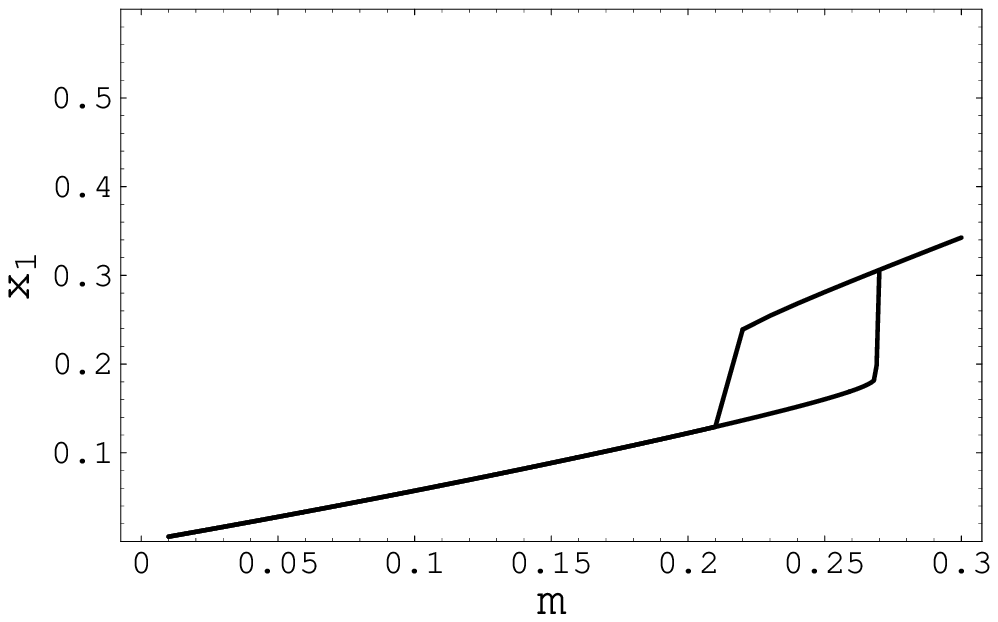}\end{center}

Figure 12. The $G_{2}/M$ transition when the gene copy number of
the MDM2 gene is one, that of the p53 gene two and the amount of DNA
damage is $A=0.17$
\end{figure}

\newpage

\section*{Acknowledgment}

B.G. is supported by the Council of Scientific and Industrial Research,
India under Sanction No. 9/15 (282)/2003 - EMR-1

\end{document}